\documentclass[12pt]{article}
\usepackage{graphicx}
\usepackage{epstopdf}
\usepackage{float}
\title{Complexity, Chaos, and the Duffing-Oscillator Model: An Analysis of Inventory Fluctuations in Markets}
\author{Varsha S. Kulkarni\\\small{\emph{School of Informatics and Computing, Indiana University, Bloomington, IN 47408, USA}}}

\date{}

\begin{document}

\maketitle
\pdfoutput=1
\textbf{Abstract}:\\
\small{Apparently random financial fluctuations often exhibit varying levels of complexity, chaos. Given limited data, predictability of such time series becomes hard to infer. While efficient methods of Lyapunov exponent computation are devised, knowledge about the process driving the dynamics greatly facilitates the complexity analysis. This paper shows that quarterly inventory changes of wheat in the global market, during 1974-2012, follow a nonlinear deterministic process. Lyapunov exponents of these fluctuations are computed using sliding time windows each of length 131 quarters. Weakly chaotic behavior alternates with non-chaotic behavior over the entire period of analysis. More importantly, in this paper, a cubic dependence of price changes on inventory changes leads to establishment of deterministic Duffing-Oscillator-Model(DOM) as a suitable candidate for examining inventory fluctuations of wheat. DOM represents the interaction of commodity production cycle with an external intervention in the market. Parameters obtained for shifting time zones by fitting the Fourier estimated time signals to DOM are able to generate responses that reproduce the true chaotic nature exhibited by the empirical signal at that time. Endowing the parameters with suitable meanings, one may infer that temporary changes in speculation reflect the pattern of inventory volatility that drives the transitions between chaotic and non-chaotic behavior.}

\section{Introduction}

Time series of financial fluctuations often tends to display complexity and chaos at varying levels. Statistical analysis of these properties has formed a major area of research inquiry among the mathematical scientists \cite{1,2,3,4}. A prior knowledge of the processes governing the dynamics of these fluctuations facilitates the correct identification of their nature. However, a typical time series such as ‘inventory’ changes in a commodity market is not indicative of the process driving it. Moreover, paucity of data hinders the accurate computation of complexity measures. Considering this, some researchers have constructed an efficient algorithm for computation of Lyapunov exponent which indicates whether the time series is chaotic or not \cite{5, 6}.\\

The present paper quantitatively examines the complexity, chaos of the quarterly inventory fluctuations of the commodity wheat in the global market for the period 1974-2012. In agricultural markets, periods of high volatility are marked by sharp rise or fall in inventories of the commodities. Inventories are stocks that accumulate due to production or deplete due to consumption. As shown already\cite{7}, demand and supply can have a more complex role in creating price panic for such changes. Highly volatile behavior of financial markets points to the existence of a complex, non-random character of financial markets. While noisy chaotic behavior of commodity markets has been examined, evidence of chaos in economic time series is weak\cite{8}. It is important to investigate the presence of nonlinearity, whether the process governing the inventory fluctuations is deterministic or stochastic \cite{9}. A deterministic process facilitates better prediction of the future by economic agents.\\

This paper attempts to establish that the Duffing-oscillator-model (DOM), studied extensively in physics research \cite{10,11,12} and sometimes applied to analyze volatility \cite{13} in commodity markets, is a credible one for the inquiry of oscillatory behavior of inventory fluctuations. The evidence here emerges from a cubic price-stock relation found for wheat during the given period. Analysis of Duffing's equation with positive damping and no external force yields stable fixed points corresponding to convergent oscillations about equilibrium. However, financial crises correspond to aberrations that arise when price instability affects formation of expectations causing destabilizing speculative behavior of traders in the market. Divergent oscillations generated by a negatively damped Duffing oscillator maybe suitable in approximating such behavior. \\ 

Further, these empirical fluctuations play an important role in market stability. An external (policy) intervention is a strategy aimed at stabilizing the inventory activity, and is typically applied at a perceived dominant frequency of the time signal. However, a change in strength of this external signal may result in transitions between chaotic and regular behaviors, or vice-versa. The Duffing's equation represents the interaction of this intervention as an external force, with the commodity production cycle \cite{14}, and how the former responds to inventory fluctuations. The parameters of the equation reflect the market situation in terms of traders’ psychology, speculation. Finally, the nature of the responses generated by the model, are compared with the true nature given by the empirical analysis of the time series.\\

The paper is organized as follows. In section 2, empirical stock fluctuations of wheat are analyzed using complexity measures. Section 3 gives the outline and derivation of the Duffing oscillator as a model for inventory fluctuations. This is followed by the analysis of the model in section 4 which includes estimation of parameters, Lyapunov exponents, and comparison with the observations. Section 5 concludes with a discussion of the key findings.\\

\section{Empirical Complexity Analysis of Inventory Fluctuations}
Figure 1 shows the quarterly changes in prices and stocks/inventory of the agricultural commodity wheat during the period 1974-2012. It reveals the pattern of oscillatory behavior of inventory changes during the period. The period of analysis covers two major food crises that occurred around 1974 and 2007. Explosion in prices around those years is evident from the figure\footnote{One of the noted reasons for this severe persistent volatility was documented as increased speculation and bad weather that affected the majority of places in the world producing wheat \cite{7}.}.

\begin{figure}[H]
\centering
\scalebox{0.75}{\includegraphics{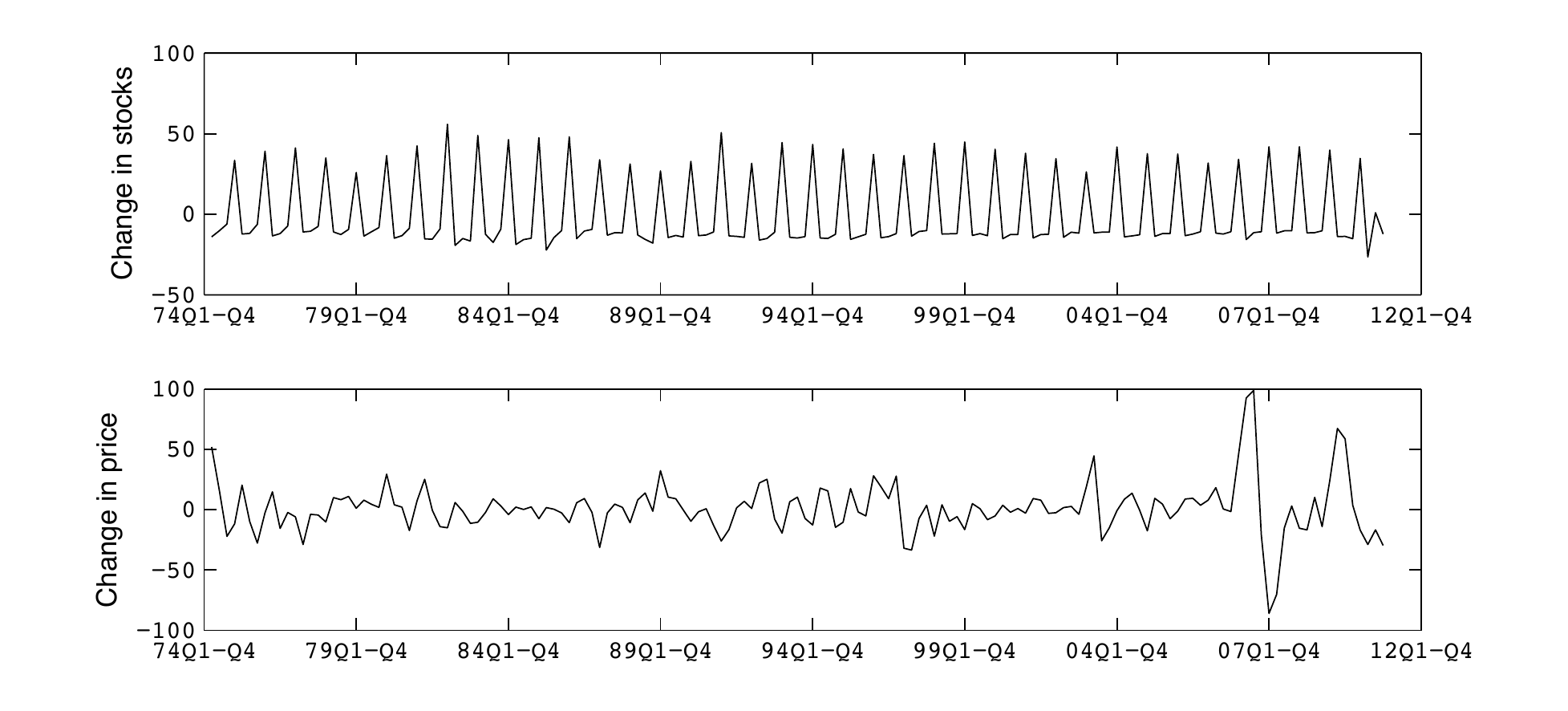}}
\caption{Temporal variation of global quarterly fluctuations of stocks (Top) and price (Bottom) of wheat for the period 1974-2012. The horizontal axis represents time in unit of quarter-year. There are 4 data points for every year and a total of 155 variations are plotted. The units of stocks and price are million ton ($mt$) and $\$/ mt$ respectively.}
\end{figure}

From Figure 2, one can see the effect of a change in stocks in one time period on its change in the next time period. Eq.(1) below gives the change in stocks $s$ in time $\Delta t$ is represented as

\begin{equation}
x(t)=s(t+\Delta t)-s(t)
\end{equation}
The plot set out in Figure 2 shows a significant negative linear relationship between $\Delta x$ and $x$ in both cases. This indicates that a sharp inventory spike at one time may lead to either a spike of similar magnitude or a dip. 

\begin{figure}[H]
\centering
\scalebox{0.75}{\includegraphics{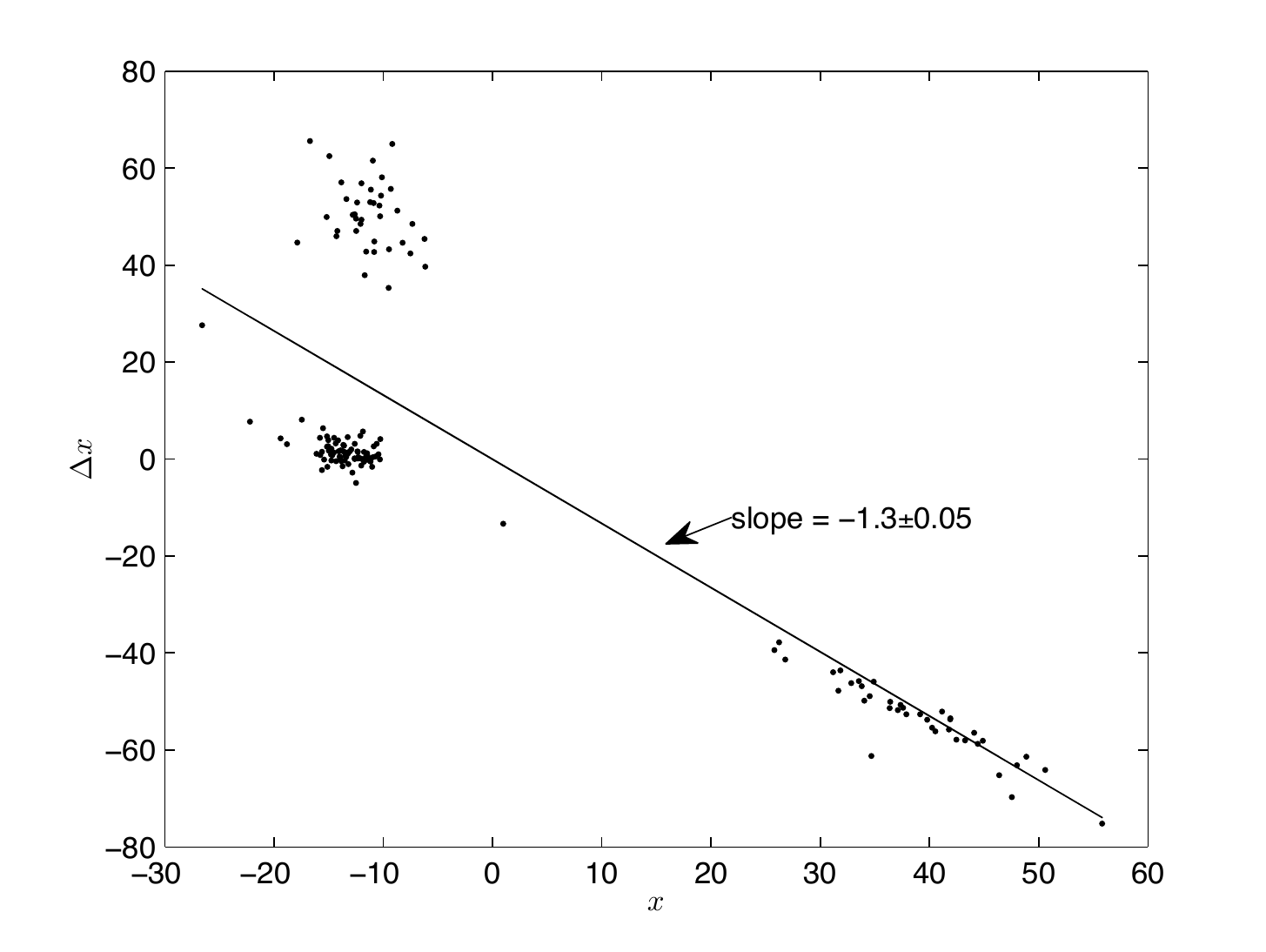}}
\caption{Change in $x$ versus $x$ at time $t$, for wheat. $\Delta t$ = 1 quarter.}
\end{figure}

\subsection{Time signal estimation}

The Fourier approximation of time signals of inventory fluctuations is expressed in the discrete form as 

\begin{equation}
x(t)= a_0 + \sum\limits_{k=1}^{N/2} a_kcos(\omega_k t)+b_ksin(\omega_k t)
\end{equation}
for a time series of length $N$.  The Fourier transform \cite{10} is $F(k)=\sum\limits_{n=1}^{N}x(t_n)e^{-i\omega_k t_n}$ for $1\leq k \leq N$. Here $\omega_k=2\pi k/N=k\omega_1$, $\omega_1$  being the fundamental angular frequency. $F(k)$ is a complex number, and used to compute Fourier coefficients $a_0$, $a_k$, and $b_k$. \\

The dominant frequencies of the time series are obtained using the periodogram\footnote{Although the spectral method works best for stationary time series, here it is an approximation for N is not very large.}, which is defined as power spectral density per unit time or $\frac{1}{N}|F|^2$. The presence of white noise in this estimation is tested for shifting time zones of 131 data points each, using the Durbin's test \cite{15}. The test employs cumulative periodogram (details are shown in appendix A) and reveals that at 0.1 level of significance, for every subperiod, some frequencies (including the dominant one) are noise free. The frequencies chosen for this analysis are the ones with greater power and significantly noise free. These are $1/4.06$ and $1/3.93$ in units of $quarter-year^{-1}$. This implies that roughly after a year, the magnitudes of fluctuations repeat atleast approximately. However, the range of this variation could make subsequent time zones different from one another.

\subsection{Detecting Nonlinearity with Correlation Dimension}

An important feature of this paper is to analyze the complexity of the stock fluctuations of  wheat and to investigate whether they can be chaotic. Chaos however requires a nonlinear dynamical process. Correlation dimension is a measure that helps to detect nonlinearity of the process generating a given time series \cite{10, 16}. It measures the minimum number of variables essential for specifying the attractor of the model dynamics. Using the standard procedure the $CD$ is estimated for a time series of length $N$ as
\begin{equation}
CD=\lim_{r\rightarrow0}\lim_{N\rightarrow\infty}\frac{dlog(C(r))}{d(logr)}
\end{equation}

The correlation sum $C(r))=\frac{2}{N(N-1)}\sum\limits_{j=1}^N\sum\limits_{j=i+1}^N\Theta(r-r_{ij})$  is computed for embedding dimension $m$ with distances $r_{ij}=\sqrt(\sum\limits_{k=0}^{m-1}(X_{i-k}-X_{j-k})^2)$. For increasing values of $m$, the slopes of straight lines on logarithmic plots of $C(r)$ versus $r$ give the values of $CD$. However, the presence of noise mars the ability of this measure to detect nonlinearity to an extent. Hence the phase randomized surrogate data testing \cite{17} is used to compare the observed values of $CD$ ($Q_0$) with those obtained from surrogate data generated by randomizing the phase information of the signal (mean= $\langle Q_s \rangle$, standard deviation=$\sigma_s$). The null hypothesis that original time series is correlated noise is rejected for a two sided test at significance level $\rho$ depending on the corresponding z-score that is, $Z=\frac{|Q_0-\langle Q_s \rangle|}{\sigma_s}$. Table 1 below summarizes the results of the test. In case of wheat, $CD$ shows a limited increase with $m$. It confirms the existence of low dimensional nonlinearity.

\begin{table}[ht]
\caption {Results of $CD$ and surrogate test conducted at $\rho=0.1$ by generating $100$ surrogates using random phase method} 
\centering 
\begin{tabular}{c c c} 
\hline\hline 
$m$ & Observed $CD$ &  Test result\\ [0.5ex]
\hline 
1 & 0.91 & Not noisy \\
2 & 1.92 & Not noisy \\
3 & 2 & Not noisy \\
4 & 2.75 & Not noisy \\
5 & 2.88 & Not noisy \\[1ex]
\hline 
\end{tabular}
\label{table1:} 
\end{table}

\subsection{Lyapunov Exponent}

Lyapunov exponent quantifies the sensitive dependence on initial conditions exhibited by the system. It hence also determines the level of predictability. A positive value of the exponent indicates chaos whereas a negative value indicates non-chaotic behavior. This subsection computes the exponent for the time series, for shifting time windows of length $N=131$ quarter-years, using a well known algorithm\cite{5}. It entails reconstruction of attractor dynamics for each time window. \\

In general for time series of length $N$, $\{t_1,t_2\ldots,t_N\}$, the state of the system at discrete time $i$ is $A_i=[t_i t_{i+J}\ldots t_{i+(m-1)J}]$ with $i=1,\ldots, M$. $J$ is the reconstruction delay and $m$ is the embedding dimension.  The reconstructed trajectory $A =[A_1 A_2 \ldots A_M]^{T}$ is an $M\times m$ matrix with $M=N-(m-1)J$. Although there is no knowledge about $m$ for the time series, the algorithm is known to be robust to change in $m$ as long as it is atleast equal to the topological dimension of the system. From the above subsection, $m=3$ seems to be a reasonable choice in this case.  $J$ is computed as the lag where the autocorrelation function drops to $1-1/e$ of the initial value. Here, it happens at $J=44$ for all time windows and so $M=43$. The largest Lyapunov exoponent $\lambda_{observed}$ is calculated as the least squares fit to the line $b(i)=\frac{\langle d_j(i)\rangle}{\Delta t}$, $d_j(i)$ denotes the distance between $j^{th}$ pair of nearest neighbors after $i$ time steps ($\Delta t$=1quarter). \\

In this case, more often than not, weak chaotic behavior is seen to alternate with non-chaotic behavior as the time window is slid forward by a quarter-year. This elucidates the variations in long term predictability. As is also apparent from the autocorrelation function, the rate of loss of predictability is relatively low.  The time series of inventory changes is marked by aberrations in the form of some large positive values that occur at almost regular intervals with negative values that are relatively moderate, and a few values that are relatively small. Further, the range of variation of the magnitudes of these positive changes cannot be ignored and may sufficiently distort the set up upon a single point shift in the time zone. These features of the data, are plausibly responsible for the observed transitions between non-chaotic and chaotic behaviors.

\section{DOM for Inventory Dynamics}

Dynamics of supply and demand tends to be complex in explaining the observed inventory volatility. What matters is the interplay with financial speculation, external forces (like policy) and other disturbances which may render the commodity system unstable. \\ 

The basic structure of the commodity production cycle \cite{14} consists of two negative feedback loops as both consumption and production adjust the inventory to the desired level. Prices fall as the inventories rise, thereby motivating a reduction in production and increase in consumption. The opposite happens when inventories fall below the appropriate level. Thus the price-stock relation is a push-pull effect wherein price change acts as a ‘restoring force’ driven by oscillations of inventory changes about the zero level. However expectations of a price rise may increase demand or consumption because producers tend to store inventory for speculation of price volatility. Thus small changes in stocks may alter the price only slightly, whereas large changes may displace it drastically in the opposite direction. This is similar to the behavior exhibited by the following nonlinear model.

\begin{equation}
\dot{p}=\alpha_1 x +\alpha_2 x^3 \hspace{0.5in} \alpha_1\alpha_2<0
\end{equation}

Here $\dot{p}$  represents the price change per year. Explanation and evidence for this relation in Eq.(5) is presented in Appendix B. Price change is considered as a nonlinear restoring force.  From the dynamics of production cycle \cite{8}, $x\approx P(p,x)-C(p,x)$, and using the empirical relation $\dot{x}\propto -x$ (Figure 2), one arrives at

\begin{equation}
\dot{x}=r(P(p,x)-C(p,x))
\end{equation}

$P$, $C$ represent the production and consumption functions respectively and $r\leq0$  is the constant of proportionality. Eqs.(4), (5) represent the coupled feedback loops of commodity oscillations.\\

The external force may be a policy intervention typically operating at the perceived dominant frequency of the empirical signal, so as to stabilize the fluctuations. If there is more than one dominant frequency in a signal, this choice may crucially affect the resultant dynamics, and even more when the signal is noisier. The force follows a rule $D(t)$ which is cyclical, of the form $asin(\omega t)$. If the force \cite{3} is initiated at time $t_0$ then $D(t_0)=0$ which requires $t_0=n\pi$; $n=0,1,2\ldots$. With this superposition Eq.(5) can be written as
\begin{equation}
\dot{x}=r\left\{P(p(t_0),x(t_0))-C(p(t_0),x(t_0))+asin(\omega(\pi-t_0))\right\}
\end{equation}

Upon rescaling the time $\tau=\pi-t_0$ and assume $n=1$ the initial conditions $p(0)$ and $x(0)$ can be determined and we have 
\begin{equation}
\dot{x}=r\left\{P(p(t),x(t))-C(p(t),x(t))+asin(\omega t)\right\} \hspace{0.5 in}t\geq \tau=0
\end{equation}

Taking the time derivative in above equation yields
\begin{equation}
\ddot{x}=r\left\{\frac{\partial{P}}{\partial{p}}\dot{p}+\frac{\partial{P}}{\partial{x}}\dot{x}-\frac{\partial{C}}{\partial{p}}\dot{p}-\frac{\partial{C}}{\partial{x}}\dot{x}+a\omega cos(\omega t)\right\}
\end{equation}

Using Eq.(4) and substituting $\delta=-r\frac{\partial{(P-C)}}{\partial{x}}, \beta=-r\alpha_{1}\frac{\partial{(P-C)}}{\partial{x}},\alpha=-r\alpha_2\frac{\partial{(P-C)}}{\partial{x}}, \gamma=ra\omega$, 
\begin{equation}
\ddot{x}+\delta\dot{x}+\beta x+\alpha x^3=\gamma cos(\omega t)
\end{equation}

Eq.(9) is the deterministic Duffing oscillator equation, an example of damped physical oscillations which may or may not be chaotic. It approximates a damped, driven inverted pendulum with torsion restoring force and describes large deflections. While it has been applied previously \cite{13} to volatile fluctuations in commodity markets, here it represents the effect of superposition of cyclic intervention (such as policies) on the cubic price-stock relationship. The parameters $\delta,\beta,\alpha,\gamma$ determine whether the fluctuations are chaotic or regular. They represent : $\delta$- extent of economic damping (due to speculation); $\beta,\alpha$- linear and nonlinear price-stock push-pull effect respectively; $\gamma$- the amplitude of the external force.

\section{Analysis}

The parameters determine the state of the system and the resultant response of an external superposition. They need to be uniquely determined for commodities as their frequency contents may be different. The Fourier time signal obtained above acts as an approximate solution of the Duffing's equation, particularly when the nonlinearity is low. As seen above, the wheat stock change signal is dominated by a pair of angular frequencies ($\omega_1,\omega_2$) or ($2\pi T_1^{-1}, 2\pi T_2^{-1}$), where $T_1,T_2$ represent the time periods in quarter-year units. The same is true for the Fourier estimation of the time signal for $x^3$. The signal for inventory fluctuations is given as
\begin{equation}
x(t)=a_0+a_1 cos(\omega_1 t)+b_1 sin(\omega_1 t)+a_2 cos(\omega_2 t)+b_2sin(\omega_2 t)
\end{equation}
And\newline
$\dot{x}=-a_1\omega_1 sin(\omega_1 t)+b_1\omega_1 cos(\omega_1 t)-a_2\omega_2 sin(\omega_2 t)+b_2 \omega_2 cos(\omega_2 t)$\newline
$\ddot{x}=-a_1\omega_1^2cos(\omega_1 t)-b_1\omega_1^2 sin(\omega_1 t)-a_2\omega_2^2cos(\omega_2 t)-b_2\omega_2^2sin(\omega_2 t)$\newline
$x^3(t)=a_0^3+Acos(\omega_1 t)+Bsin(\omega_1 t)+Ccos(\omega_2 t)+Dsin(\omega_2 t)$

These expressions are substituted in Eq.(9). Considering $\omega=\omega_1$ (the significant frequency corresponding to maximum power in the spectrum), collecting and equating the coefficients of sine and cosine terms, one gets
\begin{equation}
-a_1\omega_1^2+b_1\omega_1\delta+a_1\beta+A\alpha=\gamma
\end{equation}
\begin{equation}
-b_1\omega_1^2-a_1\omega_1\delta+b_1\beta+B\alpha=0
\end{equation}
\begin{equation}
-a_2\omega_2^2+b_2\omega_2\delta+a_2\beta+C\alpha=0
\end{equation}
\begin{equation}
-b_2\omega_2^2-a_2\omega_2\delta+b_2\beta+D\alpha=0
\end{equation}
\begin{equation}
a_0\beta+a_0^3\alpha=0
\end{equation}
Solving the set of simultaneous equations Eqs.(11-15) yields \footnote[3]{Since the number of unknowns is less than the number of equations, the least squares method gives approximate solution.} values of $\delta,\beta,\alpha,\gamma$ for $\omega=\omega_1$. 

\subsection{Lyapunov Exponent for DOM}

The parameters of Duffing’s equation are computed using sliding time windows of length 131 quarters each. For every time period, whether or not the response generated by DOM is chaotic, is determined by the Lyapunov exponent \cite{6,12,18}. Eq.(9) can be expressed as a three dimensional system $X=(x,y,t)$, with a set of first order autonomous differential equations as 
\begin{equation}
\dot{x}=y;\hspace{0.5in}
\dot{y}=-\delta y-\beta x -\alpha x^3 +\gamma cos(\omega t);\hspace{0.5in}
\vspace{0.5in}
\dot{t}=1
\end{equation}
The largest Lyapunov exponent is computed as
\begin{equation}
\lambda_{model} = \lim_{t \to \infty} \frac{1}{t}log\frac{||\Delta X(t)||}{||\Delta X(0)||}
\end{equation}
The idea involves computation of distance  between two trajectories starting infinitesimally close to each other (at time t=0), after evolving for a long time (t=t)\footnote[4]{Appendix C gives the details of the algorithm.}. A positive value $\lambda>0$ characterizes the local divergence of trajectories due to sensitive dependence on initial conditions, implying chaos, as opposed to negative value $\lambda<0$ which implies non-chaotic behavior. \\

The results are in good agreement with those obtained in section 2.3. Figure 3 compares the Lyapunov exponents generated by DOM with those obtained from the actual time series calculated above. The plot set out in Figure 4 shows the agreement between $\lambda_{observed}$ and $\lambda_{model}$ (scaled) to a reasonable approximation. A constant factor is applied to scale down the model exponents to the observed ones. This is important and specific to the commodity considered\footnote[5]{The range of time gaps $i$ used to compute $\lambda_{observed}$ in the previous section was almost fixed with very minute adjustments made in few cases to better fit the model responses while keeping the error of calculation same.}. \\

\begin{figure}[H]
\centering
\scalebox{0.75}{\includegraphics{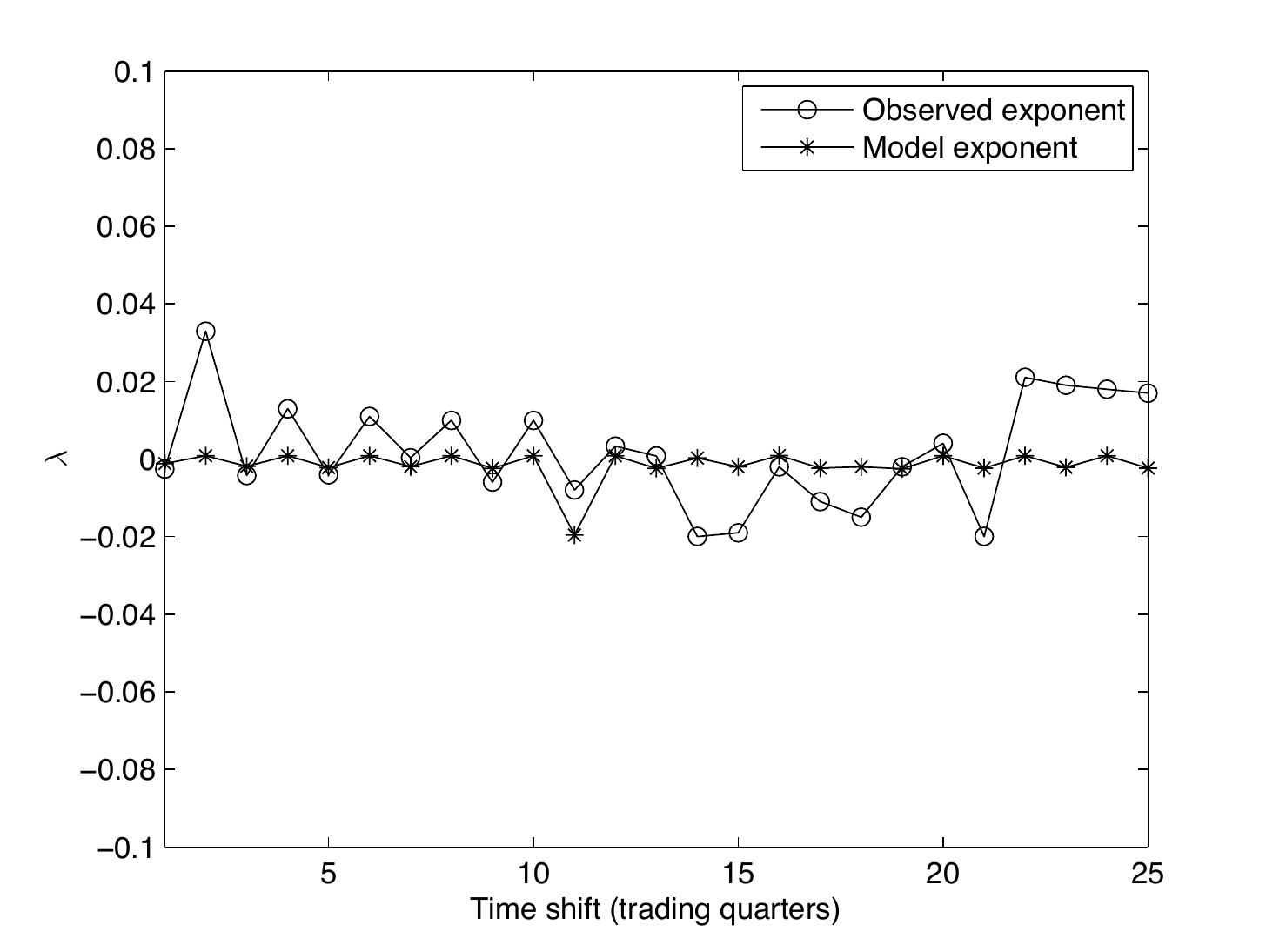}}
\caption{Plot showing variation of Lyapunov exponents computed from empirical time series directly ($\lambda_{observed}$) and DOM responses ($\lambda_{model}$) for sliding time windows of length 131 quarters each. The values of $\lambda_{model}$ are scaled down by a factor of 10.7.}
\end{figure}

\begin{figure}[H]
\centering
\scalebox{0.75}{\includegraphics{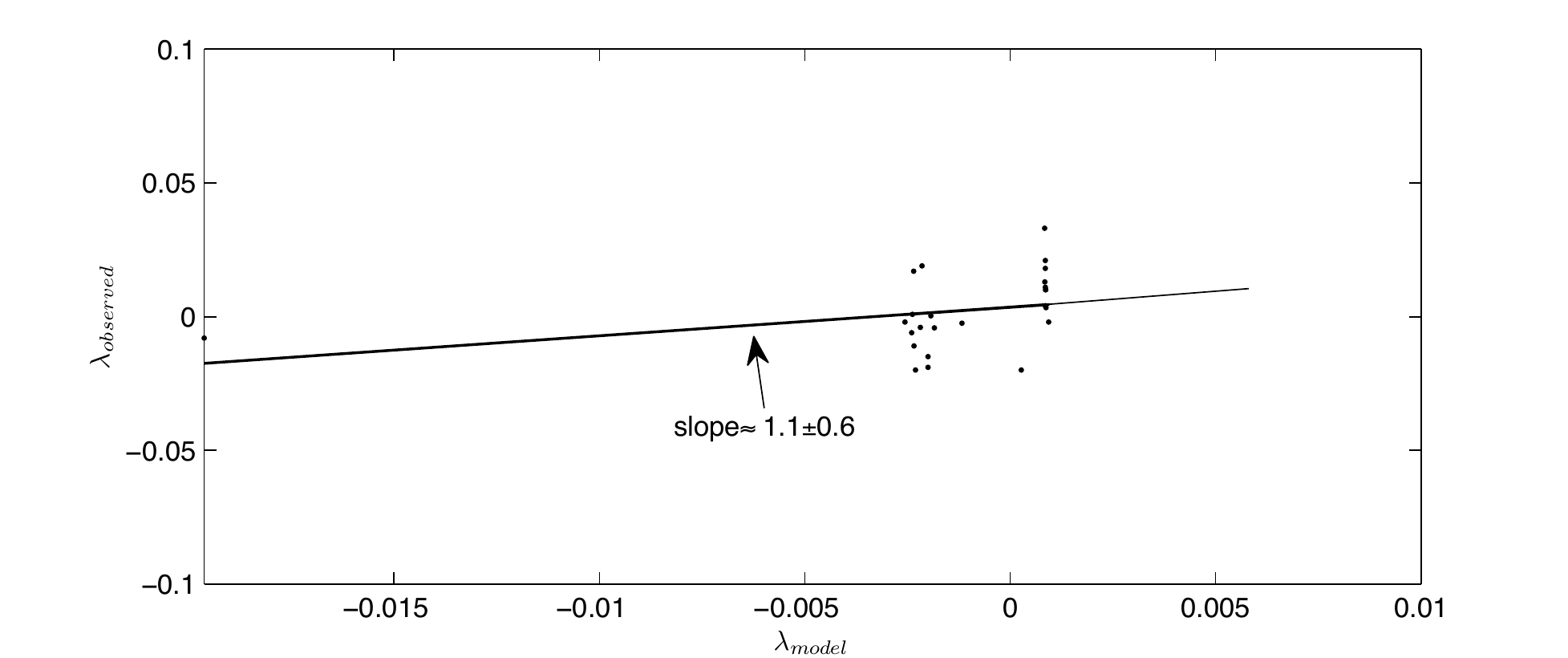}}
\caption{Plot showing agreement between $\lambda_{observed}$ versus $\lambda_{model}$ (scaled), to a reasonable approximation. The error margin could be further reduced with the use of a larger data set.}
\end{figure}

Thus, the responses generated by DOM may be considered a proxy in order to identify the nature of the complexity of the real time series. In other words, DOM explains the dynamics underlying the inventory volatility of wheat during a given time period. It must be noted that when $\delta<0$ the oscillations tend to be explosive, chaotic. Hence the role of delta as the economic damping due to speculation, is crucial. By its definition above, $\delta<0$ when  $\partial{(P-C)}/\partial{x}<0$, which implies that when inventory rises sharply, $P-C$ reduces. This is known to happen mainly because consumption/demand may rise in anticipation of price rise which motivates producers to cut back, store inventory and speculate about price volatility \cite{7,14}. \\

\section{Conclusions}

The oscillatory behavior of quarterly fluctuations of wheat inventories in the global market exhibits a complex non-random character during 1974-2012. As confirmed by the correlation dimension analysis, the dynamics of stock changes is governed by a nonlinear deterministic process. In order to investigate the chaotic nature of the time series of fluctuations, the knowledge of this process is useful. This is because the accurate computation of the Lyapunov exponent, a signature of chaos (and hence an indicator of the predictability) of the time series, is limited by insufficient amount of data. \\

The Lyapunov exponents of the wheat stock fluctuations in this paper, are computed for a given time period using an efficient algorithm established previously, based on the evolution of nearest neighbor distances. The reconstruction time delay chosen for the analysis in accordance with the loss of autocorrelation covers almost a third of the length of the time zone. Weakly chaotic behavior alternates with non-chaotic behavior in the shifting time zones over the entire period of analysis. Thus the loss of predictability in any time period occurs gradually, relatively slowly. One can examine the aberrations in the time series in the form of large positive stock changes of varying magnitudes occurring almost regularly and a few negative values of relatively small magnitudes interspersed with those. This plausibly distorts the set up and temporal evolution of distances, and causes minute but significant changes to predictability as the time window is slid forward by a single quarter. \\

The contribution of the present study lies in establishing that the deterministic DOM is able to explain the dynamics of inventory fluctuations of wheat for a given time period, with reasonable credibility. This stems mainly from the evidence of a cubic dependence of changes in price on those of stocks. Also, it facilitates a better prediction by economic agents. The strength of the external force in Duffing’s equation is determined by the fit to the Fourier estimated time signal and is indicative of the external (policy) response to the inventories at that time. Other parameters too, reflect the interaction between the commodity production cycle, and the external intervention. With suitable interpretations of the economic significance of these parameters, the dynamics of inventory changes can be quantitatively analyzed in greater detail because the properties of DOM and its responses are well known already. For instance, the response in this case is chaotic as long as the (economic) damping factor $\delta<0$. This corresponds to a situation when a rise in inventory is accompanied by a reduction in the gap between production and consumption due to speculation about price rise. \\

Further, the (scaled) Lyapunov exponents computed from the responses of the model corresponding to the time signal estimated in shifting time regimes, resemble the ones from the empirical fluctuations in those subperiods. It is plausible, therefore, that, the aberrations observed in the empirical time series which are believed to be responsible for the transitions between chaotic and non-chaotic behaviors, result from short term changes in speculation in the system/market. From this analysis, it is clear that the responses of DOM can indicate the true chaotic nature of inventory fluctuations. These responses can be further tested using other known measures of complexity. The accuracy of the agreement between the levels of complexity or chaos of the actual fluctuations and those of the model responses may be improved with larger data sets.  The hope is that this DOM analysis would also apply to inventory volatility patterns of other commodities in the market. \\

\textbf{\large{APPENDIX}}\\

\textbf{A. Durbin's test of white noise}\\
The estimation of dominant frequencies of the signal may not be noise free. Durbin's test of white noise is employed to test the presence of noise in the signal. This test statistic is\newline

$s_k=\frac{\sum\limits_{j=1}^{k}p_j}{CP}$ \hspace{0.5in} $k=1,2\ldots N/2$,\newline
for sample of size $N$ with $CP$ as the cumulative periodogram $CP=\sum\limits_{j=1}^{N/2}p_j$ and the periodogram ordinates $p_j$. For a two sided test of size $\rho$, the null hypothesis of white noise is rejected if $max_k|s_k-\frac{k}{N/2}|>c_0$ where $c_0$ is the critical value corresponding to $\rho/2$. Critical values of the test are given in \cite{15}. The test is depicted in Figure 5 and reveals mostly a noise free estimation for wheat signal. According to this analysis, two frequencies chosen are the ones outside the noise regime including the dominant frequency.

\begin{figure}[H]
\centering
\scalebox{0.75}{\includegraphics{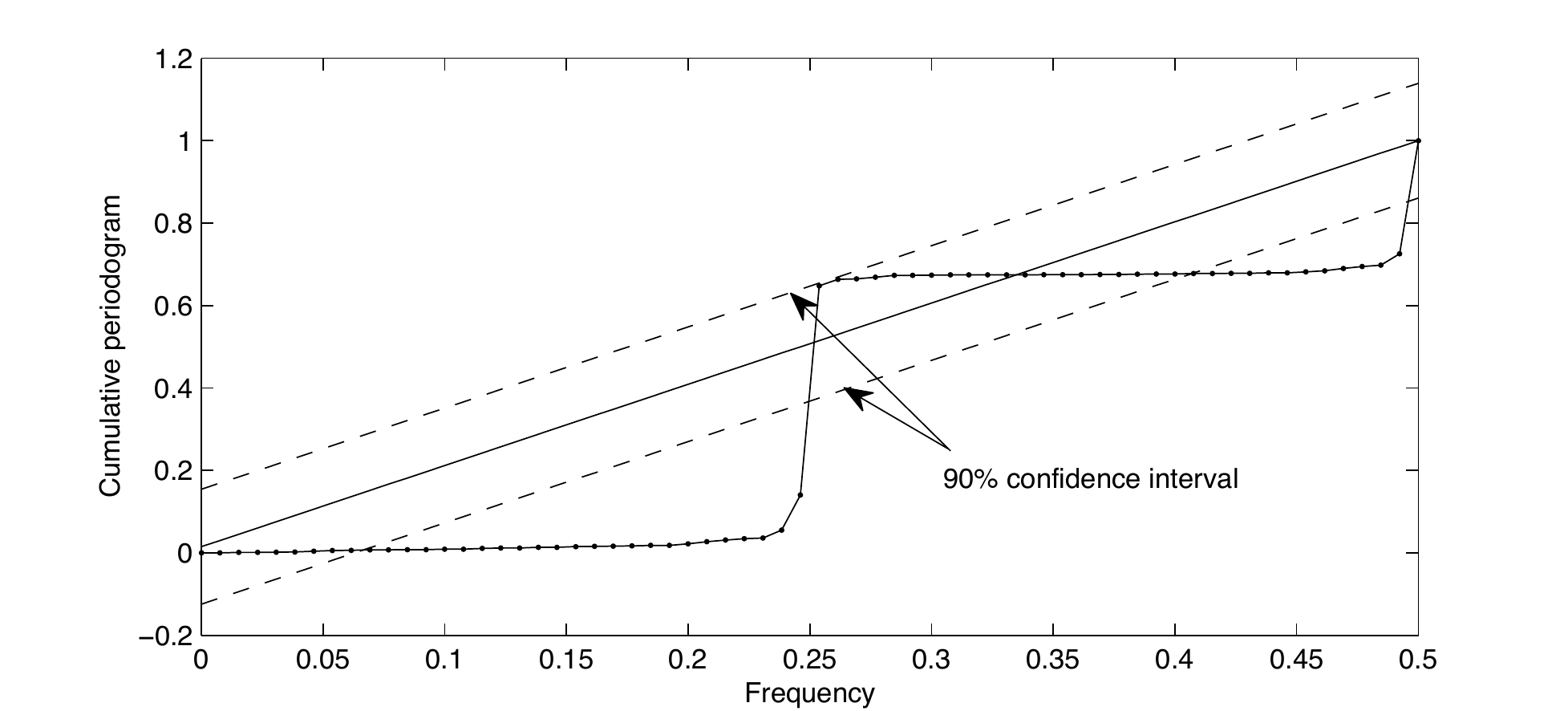}}
\caption{ Pattern of cumulative periodogram, $CP$ with frequency describes the Durbin's test of white noise for wheat. Dashed lines in both figures represent 90\% confidence intervals given by $s_k=c_0\pm k/[N/2]$ . The significance of the frequencies is indicated by the deviation of $CP$ from $45\,^{\circ}$ line (solid) corresponding to white noise. The null hypothesis of white noise is rejected for certain frequencies at $\rho=0.1$ as $CP$ crosses one of the dashed lines.}
\end{figure}

\textbf{B. Nonlinear price-stock relationship}\\

The relationship between price and stock changes is investigated. One may argue that the push-pull effect of price change on stock oscillations is analogous to restoring force effect. However, a linear relationship between price and stock changes does not adequately explain the inventory fluctuations in response to price changes. This is because small inventory changes have little effect on price change but large changes can have drastic impact on the same. The model presented in Eq.(4) exhibits such a relationship. Empirical evidence is established for the same in case of wheat. The parameters $\alpha_1$, $\alpha_2$ in Eq.(4) are estimated by ordinary least squares method and Figure 6 below shows the plot of the price change predicted by Eq.(4) with the observed price change. It provides evidence to a reasonably good approximation.
 
\begin{figure}[H]
\centering
\scalebox{0.75}{\includegraphics{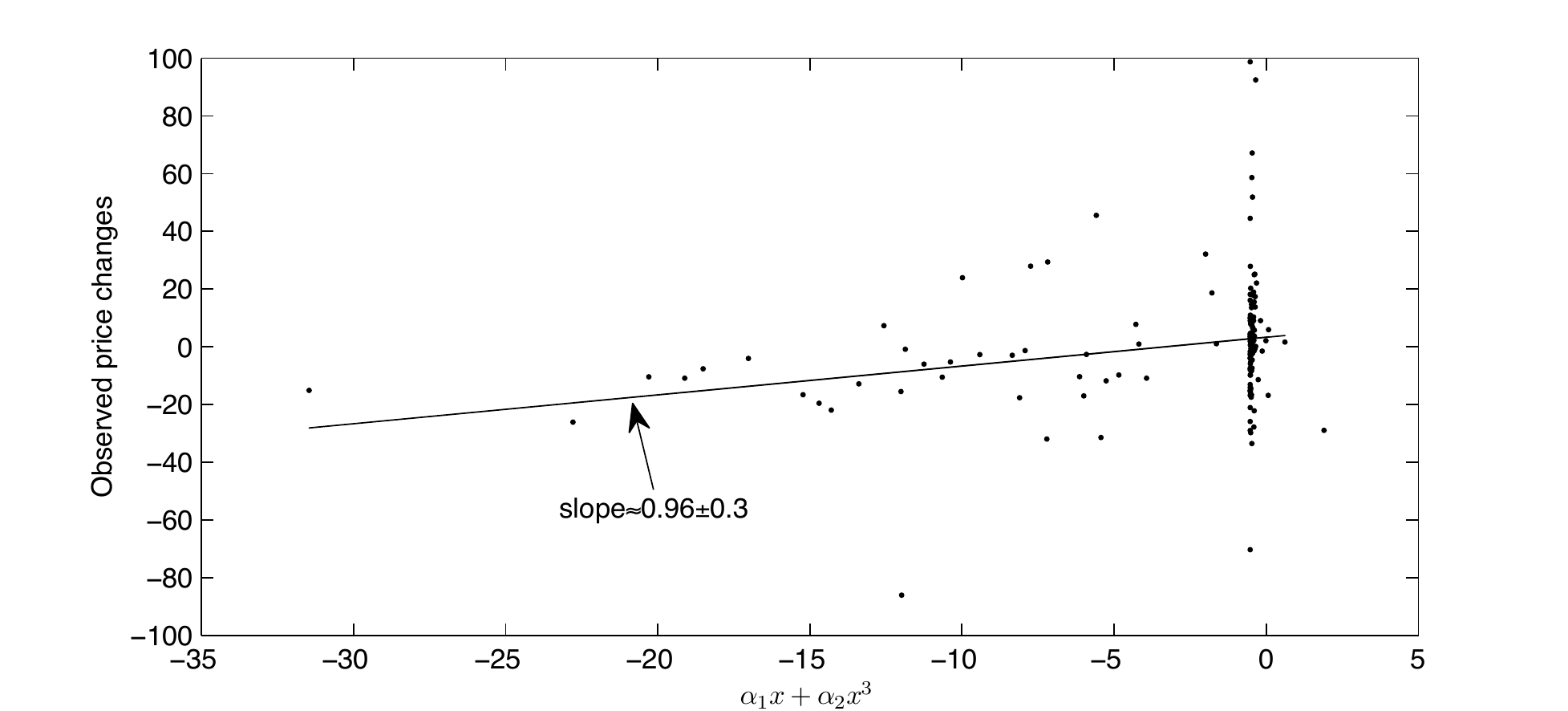}}
\caption{Plot of predicted price change modeled by Eq.(4) with the observed price change for wheat stock changes for the period of analysis. The predicted price changes are computed with significant estimates and $\alpha_1\alpha_2<0$. The slope of the best fit line in the two cases is $0.96\pm 0.3$.}
\end{figure}

This cubic relationship is an important step in establishing price change as a restoring force with nonlinear elasticity, a feature of DOM. This is because Duffing oscillator describes the motion in a quartic potential $V(x)=\frac{\beta x^2}{2}+\frac{\alpha x^4}{4}$. The signs of $\beta$, $\alpha$ determine the shape of the anharmonic potential (as double well or soft spring). In the present case, this implies that effects of linear and cubic stock change on price change together determine the motion of inventory fluctuations.\\

\textbf{C. Computation of Lyapunov Exponent}\\

Lyapunov exponent detects whether the trajectories simulated from Eq.(9), show a sensitive dependence on initial conditions, which is a hallmark of chaos. Eq.(16) with $X=(x,y,t)$ is of the form $dX/dt=f(X)$. A perturbation about the equilibrium point $X_0$ gives the variational form $dY/dt=JY$ where $J =\partial{f}/\partial{X}$ is the Jacobian matrix. The asymptotic approximation of the eigenvalues $l_i$ of $J$ at $X_0$ is used to compute the Lyapunov exponent.  In general for $n$ dimensional system with $n$ initial conditions, asymptotically,  $\lambda_i=\lim_{t \to \infty} \frac{1}{t}log|l_i|$, $i=1,2\ldots n$. The largest eigenvalue $\lambda$ so obtained is the Lypaunov exponent and characterizes the system behavior as chaotic if it is positive and non chaotic if it is negative.\\

The algorithm applied to compute $\lambda_{model}$ described in Eq.(17) is based on the standard procedure applied previously \cite{6,12,18}. The $n=3$ dimensional system, is initialized by a set of $n$ orthonormal vectors that are integrated in steps of $\Delta t$ to get $\Delta X_m$ for $m=1,2\ldots n$ over a long period of time $t=k\Delta t$. Note that in case of chaotic trajectories, the simulation needs to be terminated before the values of $\Delta X$ become very large. At each step the vectors $\Delta X_m$ are orthonormalized using the Gram Schmidt procedure as $\hat{Y_1} = \frac{\Delta X_1}{||\Delta X_1||}$ and \newline
$\hat{Y_n} = \frac{\Delta X_n - \sum\limits_{m=1}^{n-1} (\Delta X_n (\hat{Y_m}))\hat{Y_n}}{Norm_n}$, where $Norm_n=||\Delta X_n - \sum\limits_{m=1}^{n-1} (\Delta X_n (\hat{Y_m}))\hat{Y_n}||$ represents the norm of the vector at $k^{th}$ step. For the next time step, these vectors $\hat{Y_1}, \hat{Y_2}\dots \hat{Y_n}$ are taken as the new initial conditions and the process is repeated for a long time $t$. Taking the time average of log norms \newline
$\lambda_n=\frac{1}{t}\sum\limits_{i=1}^{i=k}log (Norm_n^i)$

and ordering $\lambda_1\geq\lambda_2 \ldots \geq\lambda_n$, we get the Lyapunov exponent as $\lambda_{model}=\lambda_1$.\\

\end{document}